\documentclass[useAMS,usenatbib]{mn2e}

\usepackage[letterpaper]{geometry}
\usepackage{fix2col} 

\usepackage{times}
\usepackage{booktabs} 
\usepackage{multirow} 
\usepackage{verbatim} 
\usepackage{natbib}   
\usepackage{amsmath}  
\usepackage{graphicx}

\newcommand{\prog}[1]{\textsc{\lowercase{#1}}}

\newcommand{\asym}[3]{#1$^{+#2}_{-#3}$}

\title[Timing PSR J1740$-$3052]
  {Timing the main-sequence-star binary pulsar J1740$-$3052}
\author[E. C. Madsen et al.]
  {E.~C.~Madsen,$^1$\thanks{E-mail: madsense@phas.ubc.ca}
  I.~H.~Stairs,$^1$
  M.~Kramer,$^{2,3}$
  F.~Camilo,$^4$
  G.~B.~Hobbs,$^5$
  \newauthor 
  G.~H.~Janssen,$^3$
  A.~G.~Lyne,$^3$
  R.~N.~Manchester,$^5$
  A.~Possenti,$^6$ and
  \newauthor
  B.~W.~Stappers$^3$ \\
  $^1$Department of Physics and Astronomy, University of British Columbia,
      6224 Agricultural Road, Vancouver BC V6T 1Z1, Canada\\
  $^2$Max Planck Institut f\"{u}r Radioastronomie, Auf dem H\"{u}gel 69,
      53121 Bonn, Germany\\
  $^3$Jodrell Bank Centre for Astrophysics, School of Physics and Astronomy,
      The University of Manchester, M13 9PL, UK\\
  $^4$Columbia Astrophysics Laboratory, Columbia University, New York, NY 10027, USA\\
  $^5$CSIRO Astronomy and Space Science, Australia Telescope National Facility,
      P.O. Box 76, Epping NSW 1710, Australia\\
  $^6$INAF-Osservatorio Astronomico di Cagliari, localit\`{a} Poggio dei Pini,
      Strada 54, I-09012 Capoterra, Italy}
\date{Released 2012 Xxxxx XX}

\pagerange{\pageref{firstpage}--\pageref{lastpage}} \pubyear{2012}

\def\LaTeX{L\kern-.36em\raise.3ex\hbox{a}\kern-.15em
    T\kern-.1667em\lower.7ex\hbox{E}\kern-.125emX}

\begin{document}

\label{firstpage}

\maketitle


\begin{abstract}
PSR J1740$-$3052 is a young pulsar in orbit around a companion that is most likely a B-type main-sequence star.  Since its discovery more than a decade ago, data have been taken at several frequencies with instruments at the Green Bank, Parkes, Lovell, and Westerbork telescopes.  We measure scattering timescales in the pulse profiles and dispersion measure changes as a function of binary orbital phase and present evidence that both of these vary as would be expected due to a wind from the companion star.  Using pulse arrival times that have been corrected for the observed periodic dispersion measure changes, we find a timing solution spanning 1997 November to 2011 March.  This includes measurements of the advance of periastron, $\dot{\omega}$, and the change in the projected semimajor axis of the orbit, $\dot{x}$, and sets constraints on the orbital geometry.  From these constraints, we estimate that the pulsar received a kick of at least $\sim$50 km/s at birth.  A quasi-periodic signal is present in the timing residuals with a period of 2.2$\times$ the binary orbital period.  The origin of this signal is unclear.
\end{abstract}




\section{Introduction}

Since their discovery more than 40 years ago, pulsars have proven to be incredibly useful and precise astrophysical laboratories, due in large part to our ability to time their exceptionally steady rotations.  In particular, by timing a pulsar in orbit with a binary companion, we learn not only about the pulsar, but about the companion and how it affects the pulsar's orbit and radiation.  This provides a unique method for probing properties of both the pulsar and the companion.

Of the $\sim$170 radio pulsars known to be members of binary systems, so far only four have been found whose companion is a main-sequence star.  PSR B1259$-$63 orbits a type B2e companion \citep{jml+92}, and PSR J0045$-$7319 has a B1\,V companion \citep{kjb+94}. The discovery of PSR J1638$-$4725 was reported by \citet{lfl+06} and was a result of the Parkes multibeam survey for pulsars \citep{mlc+01}; its companion spectral type has not yet been determined.  PSR J1740$-$3052, discovered by \citet{sml+01} in the same survey, orbits a companion that has been identified as most likely a late O-type or early B-type star \citep{sml+01,bbn+11}.  A coincident late-type star has been shown not to be the companion \citep{tsw+10}.  The pulsar has an orbital period of 231 days, spin period 570 ms, and characteristic age 0.35 Myr.

In this paper, we present an updated timing model for PSR J1740$-$3052 using more than a decade of data collected at several observatories. These observations are outlined in Section \ref{sec:datareduc}, along with the updated timing model and an analysis of scattering and dispersion caused by a wind from the companion.  We see that dispersion-measure (DM) variations agree very well with those predicted by a simple stellar wind velocity model.  In Section \ref{sec:discussion} we discuss secular changes in orbital parameters and the presence of a low-frequency, quasi-periodic signal in the timing residuals, and we conclude in Section \ref{sec:conclusion}.


\section{Data reduction and timing}
\label{sec:datareduc}

\subsection{Observations}
\label{sec:observations}

Data were collected between 1997 November and 2011 March on the Lovell Telescope at Jodrell Bank, UK, the Parkes telescope in New South Wales, Australia, the NRAO Green Bank Telescope in West Virginia, USA, and the Westerbork Synthesis Radio Telescope (WSRT) in the Netherlands, as listed in Table \ref{tbl:bw} along with the corresponding total bandwidths, channel bandwidths, and observing frequencies.  Data from Parkes and Jodrell Bank up to 2000 November 23 were used in the timing solution of \citet{sml+01}.

\begin{table}
\caption[Observing frequencies and bandwidths of data]{Observing frequencies and bandwidths of the J1740$-$3052 data.  Where multiple values appear in more than one field on a single line, they may be matched in the order listed.}
\label{tbl:bw}
\centering
\begin{tabular}{cccc}

\toprule

\textbf{Frequency} & \textbf{Total bw} & \textbf{Channel bw} & \textbf{Typical obs.} \\
\textbf{( MHz )}   & \textbf{( MHz )}  & \textbf{( MHz )}    & \textbf{time ( min )} \\

\midrule

\multicolumn{4}{c}{\textit{Jodrell Bank AFB (1997 Nov -- 2010 Apr)}} \\
1376               & 64, 96            & 3                   & 24 \\
1380               & 64, 96            & 3                   & 60 \\
1396               & 64                & 1                   & 24 \\
1402               & 64                & 1                   & 24 \\[1 mm]
\multicolumn{4}{c}{\textit{Jodrell Bank DFB (2009 Jan -- 2011 Mar)}} \\
1373.875           & 128               & 0.25                & 24 \\
1381.5             & 112.75            & 0.25                & 24 \\
1524               & 384, 512          & 0.5                 & 24 \\

\midrule

\multicolumn{4}{c}{\textit{Parkes AFB (1998 Oct -- 2009 Mar)}} \\
660                & 32                & 0.125               & 25 \\
1374               & 288               & 3                   & 10 \\
1390               & 256               & 0.5                 & 10 \\
1518               & 576               & 3                   & 10 \\
2350               & 288               & 3                   & 25 \\[1 mm]
\multicolumn{4}{c}{\textit{Parkes DFB (2005 Dec -- 2010 Aug)}} \\
1369               & 256               & 0.25, 0.5           & 10 \\
1433               & 256               & 0.5                 & 10 \\

\midrule

\multicolumn{4}{c}{\textit{Westerbork Synthesis Radio Telescope (1999 Oct -- 2010 Feb)}} \\
1375               & 80                & 0.156               & 30  \\
1380               & 80                & 0.156               & 20  \\

\midrule

\multicolumn{4}{c}{\textit{Green Bank BCPM (2001 Sept -- 2009 Jan)}} \\
570                & 48                & 0.5                 & 20 \\
575                & 48                & 0.5                 & 20 \\
590                & 48                & 0.5                 & 20 \\
820                & 48                & 0.5                 & 20 \\
1190               & 48                & 0.5                 & 10 \\
1400               & 96                & 1                   & 10 \\
1660               & 48                & 0.5                 & 15 \\
1780               & 48                & 0.5                 & 15 \\
2200               & 96                & 1                   & 15 \\

\bottomrule

\end{tabular}
\end{table}

At Jodrell Bank, observations are made using either an analogue filterbank (AFB) or a digital filterbank (DFB).  The AFB systems summed the orthogonal polarisation channels in hardware and used 1-bit digitisers.  The DFB system uses 8-bit digitisers and generates folded pulse profiles with 1024 bins per pulse period and full Stokes-parameter information in real time.  Total intensity profiles were formed in off-line analysis.

Observations at Parkes are also made with either an AFB or one of several DFB systems.  The sampling time on the Parkes AFB is 250~$\umu$s, and the DFB profiles had 1024 bins per period.  The PDFB data were flux-calibrated using a pulsed noise diode signal injected into the feed.

The BCPM \citep{bdz+97} at Green Bank was a filterbank with flexible channel widths and 4-bit sampling.  Profiles observed using the BCPM have a sampling time of 72~$\umu$s and are flux-calibrated using a pulsed noise diode.

WSRT data are sampled every 409.6~$\umu$s using the PuMa backend \citep{vkv02} and are not flux-calibrated.

Data obtained at every site are dedispersed assuming constant DM at values between 739 and 743~cm$^{-3}$~pc.  These variations in assumed DM do not affect the quality of the data products used here.

Individual pulses are folded at the predicted topocentric pulse period to produce pulse profiles with high signal-to-noise.  Folding is performed online, prior to calibration, in the DFB cases, and offline, after calibration, for the AFBs, BCPM, and WSRT.

Except in the case of the DFBs, data from two orthogonal polarisations are summed---during the folding process for the WSRT and the BCPM, and directly through the hardware for the AFBs.  The DFB polarisations are summed to form total-intensity profiles before obtaining pulse times-of-arrival (TOAs).

\subsection{Scattering and arrival-time fitting}
\label{sec:scattering}

To obtain TOAs, a Gaussian standard profile is first constructed from a fit to high signal-to-noise data from each different instrument, and for the various frequency ranges on each instrument, using the program \prog{bfit} \citep{kwj+94}.  Data with minimal scattering are used where possible, and \prog{bfit} is used to fit out the effects of scattering if necessary.  Each pulse profile is cross-correlated with the appropriate standard profile to produce a topocentric TOA.

Data observed at low frequencies show significant scattering, particularly at frequencies $<$1~GHz.  This is seen as a decaying exponential tail convolved with the pulse profile, as seen in Figure 1 of \citet{sml+01}.  The program \prog{Fitscatter} (written by MK) can be used to read in scattered pulse profiles along with a Gaussian standard profile and output both TOAs that account for the shifted peak induced by the convolution and scattering timescales.

We believe the scattering timescales found by the fits to be reasonable estimates.  However, the TOAs produced using \prog{Fitscatter} are less reliable than TOAs that do not account for scattering, and are not ultimately used in the full timing solution, as discussed in the next section.

\subsection{Timing analysis}

TOAs are processed using the pulsar timing software \prog{Tempo},\footnote{See http://tempo.sourceforge.net.} with barycentric corrections applied using the Jet Propulsion Laboratory DE405 solar system ephemeris \citep{sta98b} and clock corrections applied using the UTC(NIST) timescale.  The uncertainties associated with the TOAs input into \prog{Tempo} tend to be underestimates, and thus we increase the error on sets of data from different observatories and in different frequency ranges such that each set has a reduced $\chi^2$ of $\sim$1.  The 1$\sigma$ errors on parameters output by \prog{Tempo} are also doubled, providing error estimates we consider to be conservative.

\prog{Fitscatter}-produced TOAs are used for the lower-frequency data which have scattering tails that are quite long, as discussed in Section \ref{sec:scattering}, but we find that it makes no significant difference to the parameter values or reduced $\chi^2$ of the \prog{Tempo} fit (prior to error-scaling) for TOAs with frequency $>$1~GHz and only serves to reduce their correlation with higher-frequency residuals that do not show scattering.  For this reason, we do not use \prog{Fitscatter} TOAs for data at frequencies $>$1~GHz, which are the TOAs used in the full timing solution.

\subsection{Orbital dependence of DM and scattering timescale}
\label{sec:orbitaldependence}

Variations in dispersion measure (DM) near periastron were detected by \citet{sml+01} and taken as evidence for a wind from a nondegenerate companion.  The wind velocity model of \citet{ktm96} and references therein is used there to support the idea that the companion is an early B-type star rather than a late-type supergiant.  Variations in DM in this model can be expressed as
\begin{equation}
\textrm{$\Delta$DM} = \frac{1}{1.1 \times 10^{-9}} \left(\frac{\dot{M}}{v_{ratio}}\right) I~\textrm{cm$^{-3}$ pc},
\label{eqn:dmoffs}
\end{equation}
where $\dot{M}$ is the mass-loss rate of the companion in M$_{\odot}$~yr$^{-1}$ and $v_{ratio}$ is the ratio of the wind velocity at infinity to the escape velocity, expected to be in the range 1--3.  $I$ is the integral along the line of sight to the pulsar from infinity, given by
\begin{equation}
I = \int \frac{1}{\sqrt{1-R_2/r}} \frac{1}{r^2} \mathrm{d}l,
\label{eqn:integral}
\end{equation}
with $R_2$ the radius of the companion and $r$ the distance from the centre of mass of the star.

We use this model to investigate offsets in DM over the entire orbit of the binary system given orbital parameters determined through timing and integrating numerically through the wind along the line of sight to the pulsar.  

\begin{figure}
\begin{center}
\includegraphics[width=1.0\linewidth]{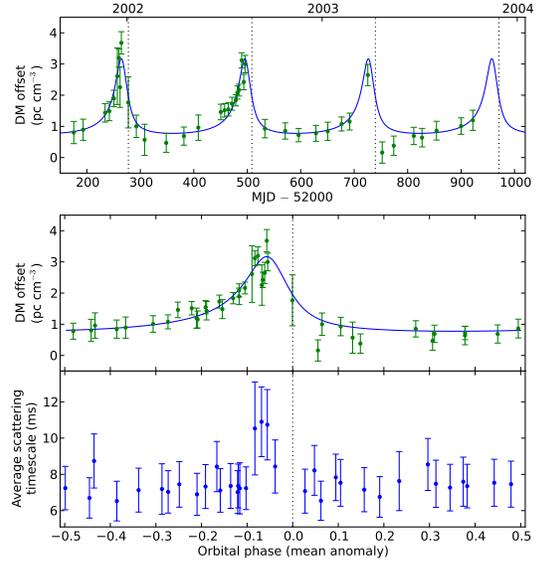}
\caption[Measured DM offsets fit to a stellar wind model, scattering plot]{Top two panels show predicted dispersion measure offsets fit to measured changes in DM over four orbits of the J1740$-$3052 system, both as a function of MJD and orbital phase.  The orbital inclination determined by this fit is $53~\pm~7$ degrees.  The offsets shown are in reference to an arbitrary `zero' value of 738.5~cm$^{-3}$~pc.  The upper horizontal axis on the top plot denotes January 1 each year.  The bottom panel shows average scattering timescales of BCPM data observed at 1190 MHz varying across orbital phase.  The data have been binned so that each point represents five measurements, except for the final point before phase zero, which represents four.  Note the increase in scattering near the phase of maximum DM, at which we expect the beam of the pulsar to traverse more of the companion's wind than elsewhere along the orbit.  In all panels, dotted vertical lines denote periastron.}
\label{fig:dmscatter}
\end{center}
\end{figure}

We calculate DM offsets by looking at four consecutive orbits (MJD 52046 to 52971) that have data spanning multiple well-separated frequencies and binning the TOAs into groups with multiple frequencies where possible, with lengths of five days or fewer.  We run \prog{Tempo} with DM arbitrarily fixed at 738.5~cm$^{-3}$~pc, allowing DM offsets to vary independently in each bin.  The \prog{Tempo} error bars are doubled, and, using the PyMC Markov Chain Monte Carlo (MCMC) algorithm \citep{phf10}, we fit these offsets to the wind-model predictions.  The orbital inclination angle is varied as a free parameter in order to find a best fit.  From the inclination angle and the assumption of a 1.4~M$_{\odot}$ pulsar, the companion's mass is calculated using the binary mass function, and this in turn is used to estimate the companion's radius, $R_2$.  This is done by interpolating over the relevant range of main-sequence masses in Table 15.8 of \citet{cox00}.  Along with the inclination angle, we vary the quantity $\dot{M}/v_{ratio}$ in Equation \ref{eqn:dmoffs} (equivalent to scaling the height of the curve in Figure \ref{fig:dmscatter}) and the baseline of the DM offsets (shifting the vertical position of the curve).  The results of this fit are shown in the first two panels of Figure \ref{fig:dmscatter}, with best fits of $53 \pm 7$~degrees, $9^{+2}_{-3} \times 10^{-10}$~M$_{\odot}$~yr$^{-1}$, and \asym{0.41}{0.15}{0.12}~cm$^{-3}$~pc for the orbital inclination, the quantity $\dot{M}/v_{ratio}$, and the DM baseline offset, respectively.  \citet{dpr+10} have measured a neutron star mass of 2~M$_{\odot}$, and so we also ran the MCMC code assuming a pulsar mass of 2.0~M$_{\odot}$.  All fit parameters were consistent within the error bars of the results listed above, and so we continue to assume a pulsar mass of 1.4~M$_{\odot}$.


It is apparent from Figure \ref{fig:dmscatter} that the model and observations agree well.  The fit has a reduced $\chi^2$ of 1.04.  Using all three output fit parameters, we numerically calculate a predicted DM offset for each TOA.  These offsets are applied to all TOAs in \prog{Tempo} to remove the effects of a varying DM from the timing residuals, even for epochs with only single-frequency observations.

We can use the fit parameters to estimate some further quantities.  Using the inclination angle obtained through this fit, and assuming a pulsar mass of 1.4~M$_{\odot}$, we estimate the mass of the companion to be \asym{20}{6}{4}~M$_{\odot}$.  This is consistent with the early main-sequence star hypothesis of \citet{sml+01}, and with the isochrones plotted in Figure 3 of \citet{bbn+11} for a main-sequence companion younger than 10~Myr, which fits with the pulsar's characteristic age of 0.35~Myr.

Allowing a $v_{ratio}$ value in the range 1--3, our measurement of $\dot{M}/v_{ratio}$ gives a mass-loss rate in the range (0.6--3)$ \times 10^{-9}$~M$_{\odot}$~yr$^{-1}$ for the companion, assuming a completely ionized wind.  While this is lower than models predict for radiatively-driven winds by at least an order of magnitude \citep[e.g.][]{vdl00,mvd+12}, some Galactic late O-type main-sequence stars have in recent years been observed to have unexpectedly low mass-loss rates in just this range \citep[e.g.][]{msh+05,mbm+09}.  This is known as the `weak-wind problem'.  The cause of these weak winds is not established, but models that initiate winds using line acceleration in addition to radiative acceleration do appear to reproduce the weak-wind problem for late O-type stars \citep{mvd+12}.  Whatever the cause, unless the wind is only partially ionized, it appears as though the companion to PSR J1740$-$3052 may experience this phenomenon.

The baseline offset is best interpreted in conjunction with the DM measured by \prog{Tempo}, and so we withhold discussion of this parameter until the next section.

Finally, we find evidence for orbital variations not only in DM, but in the scattering timescale of pulses.  In the third panel of Figure \ref{fig:dmscatter} we plot scattering timescales for 1190~MHz data as a function of orbital phase, binned to increase signal-to-noise.  It is clear that the amount of scatter increases where the pulses must traverse more of the stellar wind, which is the same phase at which the DM variations peak in the previous panel.


\subsection{Timing results}
\label{sec:timingresults}

\begin{figure*}
\begin{center}
\includegraphics[width=0.9\linewidth]{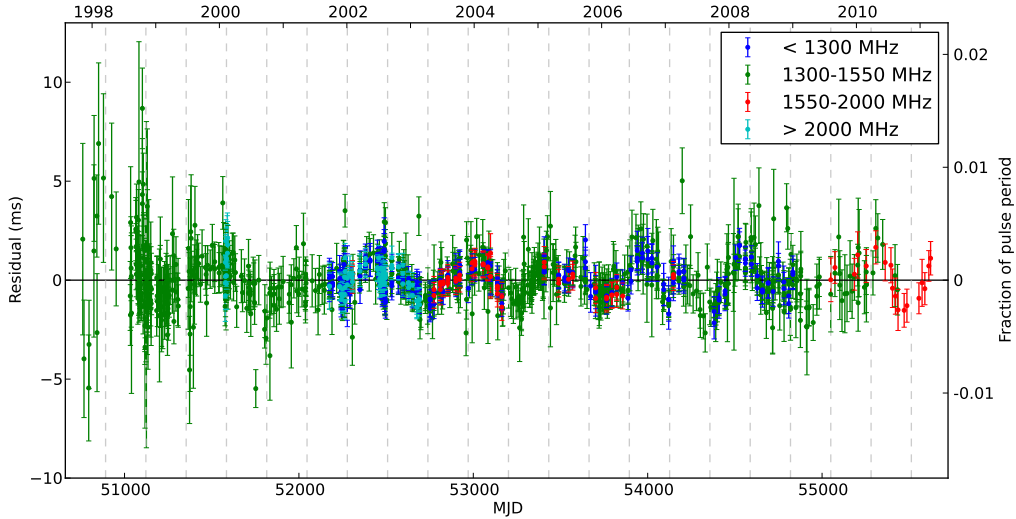}
\caption[Residuals at observing frequencies greater than 1~GHz]{Best-fit residuals using DM-corrected data observed at frequencies greater than 1~GHz.  The upper horizontal axis denotes January 1 each year, and vertical dashed lines show times at which periastron occurs.  Note the quasi-periodic signal that repeats over slightly more than two orbits of the binary system.  This is discussed further in the text.}
\label{fig:residuals}
\end{center}
\end{figure*}

We run \prog{Tempo} using the pulsar-main-sequence binary model of \citet{wex98}.  The parameters fixed and fit in \prog{Tempo} are shown in Table \ref{tbl:timing}, and the residuals for observing frequencies greater than 1 GHz are plotted in Figure \ref{fig:residuals}.  While the fit includes ten spin period derivatives (with higher-order derivatives fit as a means to reduce low-frequency timing noise), only the first two are shown.

\begin{table}
\caption[Measured and derived \prog{Tempo} fit parameters]{Measured and derived \prog{Tempo} fit parameters, with 1$\sigma$ errors reported by \prog{Tempo} doubled.  Ten spin period derivatives were used in the fit, but only the first two are reported here.  Numbers in parentheses refer to uncertainty in the last reported digits.}
\label{tbl:timing}
\centering
\begin{tabular}{lc}
\toprule
\multicolumn{2}{c}{Measured Parameters}                                                                 \\[2mm]

RA (J2000)$^a$                           & $17^{\mathrm{h}}40^{\mathrm{m}}50.\!\!^{\mathrm{s}}001$      \\
Dec. (J2000)$^a$                         & $-30^{\circ}52'04.\!\!''3$                                   \\

Period, $P$ (s)                                              & $0.570313411724(3)$                      \\
1$^{\mathrm{st}}$ Period Derivative, $\dot{P}$               & $2.5504275(95) \times 10^{-14}$          \\
2$^{\mathrm{nd}}$ Period Derivative, $\ddot{P}$ (s$^{-1}$)   & $9.1(6) \times 10^{-26}$                 \\
Epoch of Period (MJD)                                        & $53191.0$                                \\
Dispersion Measure (cm$^{-3}$ pc)$^b$                        & $738.73(8)$                              \\

Binary Model                                                 & Main Sequence Star                       \\
Orbital Period, $P_b$ (d)                                    & $231.029630(2)$                          \\
Projected Semimajor Axis, $x$ (ls)                           & $756.90794(14)$                          \\
Eccentricity, $e$                                            & $0.57887011(19)$                         \\
Longitude of Periastron, $\omega$ ($^{\circ}$)               & $178.646811(17)$                         \\
Epoch of Periastron, $T_0$ (MJD)                             & $52970.719801(11)$                       \\

Advance of Periastron, $\dot{\omega}$ ($^{\circ}$/yr)        & $0.000112(6)$                            \\
Derivative of Projected                                      & \multirow{2}{*}{$9(1) \times 10^{-12}$}  \\
\ \ \  Semimajor Axis, $\dot{x}$                                                                        \\
Second Derivative of Projected                               & \multirow{2}{*}{$< 3 \times 10^{-20}$}   \\
\ \ \  Semimajor Axis, $\ddot{x}$ (s$^{-1}$)$^c$                                                        \\
Derivative of Eccentricity, $\dot{e}$ (s$^{-1}$)$^c$         & $< 4 \times 10^{-15}$                    \\
Derivative of Orbital Period, $\dot{P_b}$$^c$                & $< 3 \times 10^{-9}$                     \\
Second Derivative of                                         & \multirow{2}{*}{$< 4 \times 10^{-23}$}   \\
\ \ \  Periastron, $\ddot{\omega}$ (rad s$^{-2}$)$^c$                                                   \\
Data Span (MJD)                                              & 50760 -- 55622                           \\

\midrule
\multicolumn{2}{c}{Derived Parameters}                                                                  \\[2mm]

Mass Function (M$_{\odot}$)                                  & 8.722676(3)                              \\
Inclination Angle, $i$ ($^{\circ}$)                          & $53(7)$                                  \\
Mass of companion, $m_2$ (M$_{\odot}$),                      & \multirow{2}{*}{\asym{20}{6}{4}}         \\
\ \ \  assuming $m_1 = 1.4$~M$_{\odot}$                                                                 \\
ISM-only Dispersion Measure                                  & \multirow{2}{*}{739.1(2)}                \\
\ \ \  (cm$^{-3}$ pc)                                                                                   \\
\bottomrule                                                                                             \\[-2mm]
\multicolumn{2}{l}{$^a$ \small{Position held fixed in \prog{Tempo}; values from \citet{bbn+11}.}}       \\
\multicolumn{2}{l}{$^b$ \small{The DM value output by \prog{Tempo} is somewhat arbitrary}}              \\
\multicolumn{2}{l}{\small{in our case, as discussed in the text.}}                                      \\
\multicolumn{2}{l}{$^c$ \small{Fit while holding all other parameters fixed at the values}}             \\
\multicolumn{2}{l}{\small{shown.}}
\end{tabular}
\end{table}

\begin{figure}
\begin{center}
\includegraphics[width=1.0\linewidth]{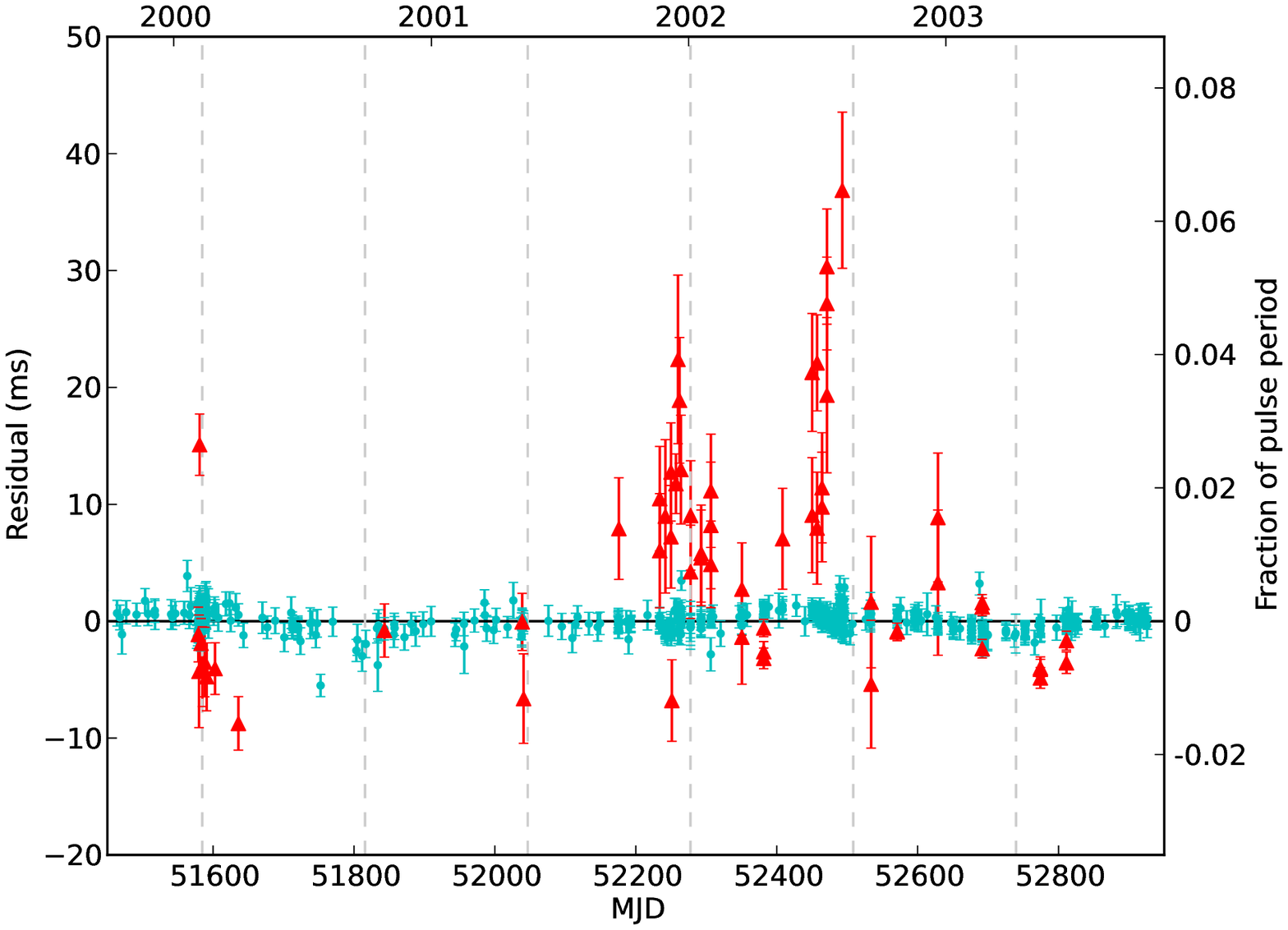}
\caption[Residuals at observing frequencies below 1~GHz]{Best-fit residuals using DM-corrected data observed at frequencies less than 1~GHz, shown as triangles, with vertical dashed lines showing times at which periastron occurs.  The other points plotted are a segment of those in Figure \ref{fig:residuals}, also corrected for DM variations.  The low-frequency TOAs are produced by \prog{Fitscatter} to account for scattering due to stellar wind and the interstellar medium.  The amount of delay in these low-frequency residuals is considerably more than we would expect even from our DM-variation model.}
\label{fig:res_lowfreq}
\end{center}
\end{figure}

The lower-frequency data do not appear to follow the $f^{-2}$ variation in arrival times between data at different frequencies as the $>$1~GHz data do, and so our timing solution excludes these data.  Residuals that include these points are shown in Figure \ref{fig:res_lowfreq} to illustrate the poor fit.  These TOAs account for timing offsets due to their large scattering timescales, as described in section \ref{sec:scattering}.

\prog{Fitscatter} is producing TOAs in these low-frequency bands that are late compared to the rest of our timing solution by up to $\sim$7\% of the pulse period.  Profiles in this frequency range are very noisy, making it difficult to investigate the source of this delay.  Perhaps the model used for scattering in the interstellar medium cannot be applied directly to scattering caused by a stellar wind.  Some pulsars also show significant variations in profile shape between widely-spaced frequencies \citep[e.g.][]{hsh+12,pw92}; this could affect timing at lower frequencies if present and undetected.

\begin{figure}
\begin{center}
\includegraphics[width=1.0\linewidth]{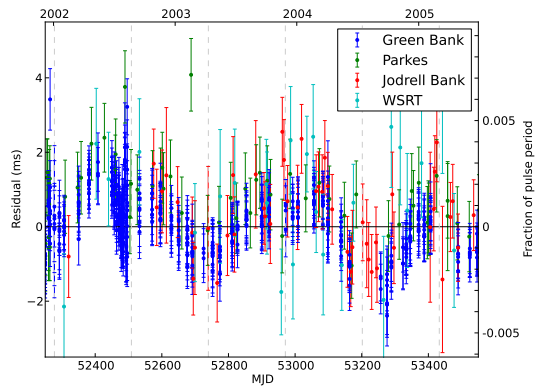}
\caption[Residuals for fit with four frequency derivatives]{A segment of the residuals after fitting only four frequency derivatives rather than ten, with colours representing different telescopes.  It is clear that the 2.2$P_b$ signal is present at all sites, and is not merely an artefact caused by fitting out a higher-order polynomial.  Axis labels are the same as in Figure \ref{fig:residuals}, and vertical dashed lines show times at which periastron occurs.}
\label{fig:res_f4}
\end{center}
\end{figure}

The DM value output by \prog{Tempo} represents the baseline from which our orbital-phase-dependent DM offsets are measured, and this allows us to estimate how much of the DM is due to the interstellar medium (ISM) rather than the companion's wind.  We expect it to be close to the arbitrary value of 738.5~cm$^{-3}$~pc we used as a baseline when determining the offsets, but there is no reason it should be the same, as we are now applying a model DM offset to our entire set of TOAs.  Indeed, we measure 738.73(8)~cm$^{-3}$~pc.  As stated in the previous section, the theoretical curve of Figure \ref{fig:dmscatter} was shifted up by \asym{0.41}{0.15}{0.12}~cm$^{-3}$~pc to fit the measured DM offsets, and so our offsets are too high by this amount, or equivalently, our baseline DM value is too low by this amount.  Adding this value, therefore, to the DM output by \prog{Tempo}, we obtain a value of 739.1(2)~cm$^{-3}$~pc, which is an estimate of the DM resulting only from the ISM.  It should be noted that the companion wind's contribution should be included when observing PSR J1740$-$3052, ideally with the orbital phase at the time of observation taken into account to dedisperse as accurately as possible.

It is clear from Figure \ref{fig:residuals} that there remain some unmodelled systematics in our timing residuals, particularly a prominent quasi-periodic signal that is present in data acquired at various frequencies and observatories.  Figure \ref{fig:res_f4} shows a segment of the residuals for a fit using only four spin derivatives, rather than ten, to emphasize that we do not think these remaining systematics are merely an artefact caused by fitting out a high-order polynomial.  Using a Lomb-Scargle periodogram \citep[][\S 13.8]{sca82,ptvf07}, we measure the period of this signal to be \asym{1.40}{0.07}{0.10} years, or \asym{2.22}{0.11}{0.16} orbital periods, where the error bars are taken as the change in frequency (subsequently converted to period) required to reach half the peak value on the periodogram.  This signal will be discussed further below.

\section{Discussion}
\label{sec:discussion}

\subsection{Secular changes in orbital parameters}
\label{sec:secularchanges}

The value measured for the advance of periastron here, $\dot{\omega}~=~0.\!\!^{\circ}000112(6)$~yr$^{-1}$, is smaller than the value $0.\!\!^{\circ}00021(7)$~yr$^{-1}$ measured by \citet{sml+01}, though they are consistent within twice the latter's 1$\sigma$ error bars.  Following the calculation in Equation 2 of that paper, but with $m_1~=~1.4~$M$_{\odot}$ and $m_2~=~20~$M$_{\odot}$, general relativity predicts an advance of $\dot{\omega}~=~0.\!\!^{\circ}00026$~yr$^{-1}$.  This is approximately double the value measured here, suggesting that there is a counteracting classical contribution due to a mass quadrupole.  Conversely, if we assume our measured $\dot{\omega}$ to be due entirely to general relativity, we can determine the total mass of the system using this value along with the measured orbital period and eccentricity.  This provides a 3$\sigma$ upper limit of 7.3~M$_{\odot}$, which is considerably smaller than the 8.7~M$_{\odot}$ binary mass function.  Thus, without a counteracting classical contribution, we would require a \emph{negative} pulsar mass.

\begin{figure}
\begin{center}
\includegraphics[width=1.0\linewidth]{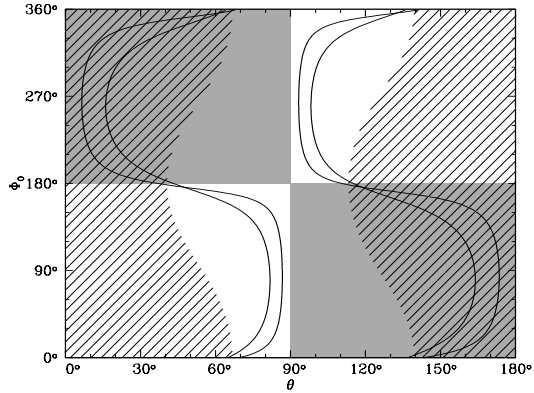}
\caption[$\Phi_0$--$\theta$ constraints diagram]{Constraints on the angle of precession of the orbital plane, $\Phi_0$, and the angle between the main-sequence star's spin and orbital angular momenta, $\theta$, given the masses of the two stars and the inclination angle of the system.  The grey regions are excluded because $\dot{x}$ is positive and the hatched regions are excluded because the classical contribution to $\dot{\omega}$ is negative.  Because the inclination angle is by far the dominant source of error, the solid lines are drawn assuming the nominal measured values of $\dot{x}$ and $\dot{\omega}$.  If the spin and orbital angular momenta are closer to being aligned than anti-aligned, we find that $40^{\circ} < \theta < 87^{\circ}$, with $\theta$ close to $80^{\circ}$ for most values of $\Phi_0$.  Otherwise, we may subtract these angles from $180^{\circ}$.  After Wex 1998, Figure 11.}
\label{fig:wexdiagram}
\end{center}
\end{figure}

\citet{sml+01} show that for a B-type main sequence star companion, a mass quadrupole due to tidal effects is negligible compared to a spin-induced quadrupole.  Following their methods, we find that the classical contribution to $\dot{\omega}$ is approximately $-0.\!\!^{\circ}00015$~yr$^{-1}$.  Using this and our measurement of the rate of change of the projected semi-major axis $\dot{x}$, for which there was previously only an upper limit, we find a greater angle between spin and orbital angular momentum vectors in the companion than was previously estimated.  As shown in Figure \ref{fig:wexdiagram}, which uses the model of \citet{wex98}, the companion's spin momentum differs by 40$^{\circ}$--87$^{\circ}$ from being either aligned or anti-aligned with the orbital angular momentum of the binary system.  Such a large relative tilt would likely have been induced by an asymmetry or `kick' in the supernova explosion that created the pulsar \citep{kbm+96}.  Making the assumption that the angular momenta in the pre-supernova system were aligned and that the orbit was circular \citep{kal96}, we find the kick velocity was at least 50~km/s.

A known alternative mechanism for producing secular changes in $\dot{x}$ and $\dot{\omega}$ arises simply from the change in viewing geometry due to the proper motion of the system \citep[e.g.][]{kop96}.  Proper motion has not been measured in our system, but we can easily see that this effect may be disregarded.  A transverse speed of 500~km/s, absurdly high for such a massive system, leads to a motion of about 10~mas/yr at the estimated distance of 11~kpc.  This would contribute at most $\sim\!4 \times 10^{-6}$~deg~yr$^{-1}$ to $\dot{\omega}$ and $\sim\!9 \times 10^{-13}$ to $\dot{x}$.  Comparing with our measured values in Table \ref{tbl:timing} or with the classical contribution to $\dot{\omega}$ above, we see that this effect falls short of our measurements by an order of magnitude or more in each case.  Thus, we consider it no further.

The distribution of angles shown in Figure \ref{fig:wexdiagram} leads to a 3$\sigma$ lower limit of $3.2 \times 10^{-7}$~AU$^2$ on the quadrupole moment $q$ of the companion.  This is determined by allowing $\dot{\omega}$, $\dot{x}$, and the inclination angle to vary by 3$\sigma$ and finding the allowed combination of angles $\theta$ and $\Phi_0$ that minimizes $q$ in the equations of \citet{wex98}.  This cannot place an upper limit on $q$, but a Monte Carlo sampling of the parameter space gives a median value that is only about twice the lower limit at $6.7 \times 10^{-7}$~AU$^2$, assuming a uniform distribution over the orbital-plane precession angle $\Phi_0$.  Taking this median $q$, companion mass 20~M$_{\odot}$ and a stellar radius of 8~R$_{\odot}$  determined by interpolating a standard table of main-sequence masses and radii \citep{cox00}, we can use Equation 6 of \citet{sml+01} to estimate roughly the stellar spin period $P_S$.  We find $P_S \approx 2 \times 10^{5}$~s, which remains consistent with the value assumed by \citet{sml+01}.

\subsection{Low-frequency residual signal}
\label{sec:lowfreqsignal}

The source of the 2.2$P_b$ signal in the timing residuals is not known.  For the sake of interest, we imagine a third body orbiting the pulsar-main-sequence binary system with this period.  As a simple approximation, we treat the binary system as a single body of mass $20 + 1.4 = 21.4$~M$_{\odot}$ located at its centre of mass in a circular orbit with a planet.  We also assume the inclination of this orbit to be the same as that for the binary system, which we fixed at 53$^{\circ}$.  We find that in such an approximation, given the uncertainties in inclination angle, main-sequence star mass, and the amplitude of the signal (which we estimate to be anywhere from slightly less than 1~ms to 2~ms), a planet would be 2--8 Earth masses, orbiting at a radius of 3--4~AU.

Given the inclination angle, the semimajor axis of the inner binary is $1.9\pm0.2$~AU, with a maximum separation of $3.0\pm0.3$~AU.  A circumbinary planet thus seems unlikely at such a short distance, making this rough approximation unappealing.  \citet{db11} have run simulations to test the stability of circumbinary orbits.  The results of those simulations suggest that a system with the eccentricity and mass ratio of J1740$-$3052 could not host a stable circumbinary orbit closer than roughly 5 times the semimajor axis of the binary (e.g.\ Figure 14 of that paper).

Furthermore, a planet orbiting only the pulsar with this period would be at roughly the same distance as the companion star, which is untenable.

A more satisfying explanation would be one that does not invoke an additional orbiting body.  If it were some consequence of the companion's spin, we would expect a much shorter period, as main-sequence stars rotate with periods on the order of days, not years \citep[e.g.][Table 15.8]{cox00}.  In particular, in Section \ref{sec:secularchanges} above, we estimated the spin period of the companion to be only a few days. 

\citet{as94} explain such a `superorbital' period in eclipsing pulsar binary PSR B1957$+$20 by variations in the quadrupole moment induced by tidal activity.  The far greater separation and companion mass in the PSR J1740$-$3052 system make this seem a very unlikely mechanism for the observed variations, as well as our assumption in the previous section that tidal effects are small in this system.  The orbit of the pulsar being tilted at least 40$^{\circ}$ and perhaps close to 90$^{\circ}$ from the spin of the companion would further reduce the effectiveness of such a mechanism.  

If the companion is a Be star, as in the case of PSR B1259$-$63 \citep{jml+92}, its circumstellar disk may be tilted due to the misalignment of the star's spin and orbital angular momenta, and might thus be expected to precess \citep[e.g.][]{mptl11}.  A precessing disk would introduce a new source of periodicity into the system that might explain the observed variations.  \citet{mptl11} suggest that the B-star binary PSR J0045$-$7319 should show oscillations on a timescale of about a year if a disk is periodically ejected, and the same might apply to J1740$-$3052.  Evidence of a disk around the companion star (such as hydrogen emission lines) would put constraints on such a scenario in our system.

\section{Conclusion}
\label{sec:conclusion}

We have updated the \citet{sml+01} timing model for PSR J1740$-$3052 using data acquired on several instruments.  From the timing results, we see orbital-phase variations in the scattering and dispersion of the pulsar's radio beam that are clear indicators of a stellar wind coming from its massive companion, lending further support to the pulsar-main-sequence binary picture for this system.  Furthermore, the dispersion-measure variations fit very well to a simple radiative stellar wind model for the charged-particle column density.  The mass-loss rate derived from this fit is lower than predicted by models, but matches the rates observed in some late O-type main-sequence stars found in the Galaxy in recent years.  This weak-wind problem remains an area of active current research.  If the companion of PSR J1740$-$3052 is indeed such a star, then its spectral class is presumably late O-type rather than early B-type.

With our new timing solution, we are able to measure the derivative of the projected semi-major axis $\dot{x}$, which leads to new constraints on the geometry of the binary system---the pulsar's orbit appears to be tilted substantially from being aligned (or anti-aligned) with the companion's spin.  This tilt of the orbital plane was likely caused by asymmetries in the supernova explosion that produced the pulsar, implying a kick of at least $\sim$50~km/s.

The origin of the quasi-periodic 2.2$P_b$ signal seen in the timing residuals of Figure \ref{fig:residuals} is a mystery.  \citet{bbn+11} suggest that spectroscopic observations of the companion could be made with the aid of adaptive optics, further constraining the orbital parameters and spectral class of the star.  Such constraints might narrow down the possible mechanisms for producing this signal, particularly if evidence is found for a circumstellar disk.

Finally, considering the winds of both the pulsar and its main-sequence companion, we might expect a shock front between the two to produce emission at X-ray and gamma-ray energies in this binary system.  High-energy emission has been observed in association with the pulsar-main-sequence binary B1259$-$63 \citep{tak94} and the millisecond pulsar B1957$+$20 \citep{sgk+03}, in both cases attributed to a shocked pulsar wind.  \citet{sml+01} examined archival X-ray observations in the field of PSR J1740$-$3052 and found no significant emission, but were unable to rule out the existence of shock emission.

PSR~J1740$-$3052 is one of just a few systems that represents a particular early stage in the evolution of binary systems containing neutron stars.  Most such binaries presumably begin as a pair of massive main-sequence stars and pass through a relatively brief phase during which one has evolved to form a neutron star, while the other still burns hydrogen in its core.  Not only does this strengthen our models of binary evolution; as seen here, it provides a unique means of probing the wind and gravitational properties of a main-sequence star.

\section*{Acknowledgments}

The Green Bank Telescope is operated by the National Radio Astronomy Observatory, a facility of the National Science Foundation operated under cooperative agreement by Associated Universities, Inc.  The Parkes Observatory is part of the Australia Telescope, which is funded by the Commonwealth of Australia for operation as a National Facility managed by CSIRO.  The Westerbork Synthesis Radio Telescope is operated by ASTRON (Netherlands Institute for Radio Astronomy) with support from the Netherlands Organisation for Scientific Research (NWO).  Pulsar research at UBC is supported by an NSERC Discovery Grant.  We thank Robert Ferdman for help in acquiring some of the GBT data and several people for help with Parkes observing.



\begin{thebibliography}{32}
\expandafter\ifx\csname natexlab\endcsname\relax\def\natexlab#1{#1}\fi

\bibitem[{Applegate \& Shaham(1994)}]{as94}
Applegate J.~H., Shaham J., 1994, ApJ, 436, 312

\bibitem[{Backer {et~al}\mbox{.}(1997)Backer, Dexter, Zepka, D., Wertheimer,
  Ray, \& Foster}]{bdz+97}
Backer D.~C., Dexter M.~R., Zepka A., D. N., Wertheimer D.~J., Ray P.~S.,
  Foster R.~S., 1997, PASP, 109, 61

\bibitem[{Bassa {et~al}\mbox{.}(2011)Bassa, Brisken, Nelemans, Stairs,
  Stappers, \& Kramer}]{bbn+11}
Bassa C.~G., Brisken W.~F., Nelemans G., Stairs I.~H., Stappers B.~W., Kramer
  M., 2011, MNRAS, 412, L63

\bibitem[{{Cox}(2000)}]{cox00}
{Cox} A.~N., ed., 2000, Allen's astrophysical quantities, 4th ed. AIP Press;
  Springer, New York

\bibitem[{{Demorest} {et~al}\mbox{.}(2010){Demorest}, {Pennucci}, {Ransom},
  {Roberts}, \& {Hessels}}]{dpr+10}
{Demorest} P.~B., {Pennucci} T., {Ransom} S.~M., {Roberts} M.~S.~E., {Hessels}
  J.~W.~T., 2010, Nature, 467, 1081

\bibitem[{Doolin \& Blundell(2011)}]{db11}
Doolin S., Blundell K.~M., 2011, MNRAS, 418, 2656

\bibitem[{{Hassall} {et~al}\mbox{.}(2012){Hassall}, {Stappers}, {Hessels},
  {Kramer}, {Alexov}, {Anderson}, {Coenen}, {Karastergiou}, {Keane},
  {Kondratiev}, {Lazaridis}, {van Leeuwen}, {Noutsos}, {Serylak}, {Sobey},
  {Verbiest}, {Weltevrede}, {Zagkouris}, {Fender}, {Wijers}, {Bahren}, {Bell},
  {Broderick}, {Corbel}, {Daw}, {Dhillon}, {Eisloffel}, {Falcke},
  {Griessmeier}, {Jonker}, {Law}, {Markoff}, {Miller-Jones}, {Osten}, {Rol},
  {Scaife}, {Scheers}, {Schellart}, {Spreeuw}, {Swinbank}, {ter Veen}, {Wise},
  {Wijnands}, {Wucknitz}, {Zarka}, {Asgekar}, {Bell}, {Bentum}, {Bernardi},
  {Best}, {Bonafede}, {Boonstra}, {Brentjens}, {Brouw}, {Bruggen}, {Butcher},
  {Ciardi}, {Garrett}, {Gerbers}, {Gunst}, {van Haarlem}, {Heald}, {Hoeft},
  {Holties}, {de Jong}, {Koopmans}, {Kuniyoshi}, {Kuper}, {Loose}, {Maat},
  {Masters}, {McKean}, {Meulman}, {Mevius}, {Munk}, {Noordam}, {Orru}, {Paas},
  {Pandey-Pommier}, {Pandey}, {Pizzo}, {Polatidis}, {Reich}, {Rottgering},
  {Sluman}, {Steinmetz}, {Sterks}, {Tagger}, {Tang}, {Tasse}, {Vermeulen}, {van
  Weeren}, {Wijnholds}, \& {Yatawatta}}]{hsh+12}
{Hassall} T.~E. {et~al.}, 2012, A\&A, submitted, arXiv:1204.3864

\bibitem[{Johnston {et~al}\mbox{.}(1992)Johnston, Manchester, Lyne, Bailes,
  Kaspi, Qiao, \& D'Amico}]{jml+92}
Johnston S., Manchester R.~N., Lyne A.~G., Bailes M., Kaspi V.~M., Qiao G.,
  D'Amico N., 1992, ApJ, 387, L37

\bibitem[{Kalogera(1996)}]{kal96}
Kalogera V., 1996, ApJ, 471, 352

\bibitem[{Kaspi {et~al}\mbox{.}(1996{\natexlab{a}})Kaspi, Bailes, Manchester,
  Stappers, \& Bell}]{kbm+96}
Kaspi V.~M., Bailes M., Manchester R.~N., Stappers B.~W., Bell J.~F.,
  1996{\natexlab{a}}, Nature, 381, 584

\bibitem[{Kaspi {et~al}\mbox{.}(1994)Kaspi, Johnston, Bell, Manchester, Bailes,
  Bessell, Lyne, \& D'Amico}]{kjb+94}
Kaspi V.~M., Johnston S., Bell J.~F., Manchester R.~N., Bailes M., Bessell M.,
  Lyne A.~G., D'Amico N., 1994, ApJ, 423, L43

\bibitem[{Kaspi {et~al}\mbox{.}(1996{\natexlab{b}})Kaspi, Tauris, \&
  Manchester}]{ktm96}
Kaspi V.~M., Tauris T., Manchester R.~N., 1996{\natexlab{b}}, ApJ, 459, 717

\bibitem[{Kopeikin(1996)}]{kop96}
Kopeikin S.~M., 1996, ApJ, 467, L93

\bibitem[{Kramer {et~al}\mbox{.}(1994)Kramer, Wielebinski, Jessner, Gil, \&
  Seiradakis}]{kwj+94}
Kramer M., Wielebinski R., Jessner A., Gil J.~A., Seiradakis J.~H., 1994,
  A\&AS, 107, 515

\bibitem[{{Lorimer} {et~al}\mbox{.}(2006){Lorimer}, {Faulkner}, {Lyne},
  {Manchester}, {Kramer}, {McLaughlin}, {Hobbs}, {Possenti}, {Stairs},
  {Camilo}, {Burgay}, {D'Amico}, {Corongiu}, \& {Crawford}}]{lfl+06}
{Lorimer} D.~R. {et~al.}, 2006, MNRAS, 372, 777

\bibitem[{Manchester {et~al}\mbox{.}(2001)Manchester, Lyne, Camilo, Bell,
  Kaspi, D'Amico, McKay, Crawford, Stairs, Possenti, Morris, \&
  Sheppard}]{mlc+01}
Manchester R.~N. {et~al.}, 2001, MNRAS, 328, 17

\bibitem[{Marcolino {et~al}\mbox{.}(2009)Marcolino, Bouret, Martins, Hillier,
  Lanz, \& Escolano}]{mbm+09}
Marcolino W.~L.~F., Bouret J.-C., Martins F., Hillier D.~J., Lanz T., Escolano
  C., 2009, A\&A, 498, 837

\bibitem[{{Martin} {et~al}\mbox{.}(2011){Martin}, {Pringle}, {Tout}, \&
  {Lubow}}]{mptl11}
{Martin} R.~G., {Pringle} J.~E., {Tout} C.~A., {Lubow} S.~H., 2011, MNRAS, 416,
  2827

\bibitem[{{Motch} {et~al}\mbox{.}(2005){Motch}, {Sekiguchi}, {Haberl},
  {Zavlin}, {Schwope}, \& {Pakull}}]{msh+05}
{Motch} C., {Sekiguchi} K., {Haberl} F., {Zavlin} V.~E., {Schwope} A., {Pakull}
  M.~W., 2005, A\&A, 429, 257

\bibitem[{Muijres {et~al}\mbox{.}(2012)Muijres, Vink, de~Koter, M{\"u}ller, \&
  Langer}]{mvd+12}
Muijres L.~E., Vink J.~S., de~Koter A., M{\"u}ller P.~E., Langer N., 2012,
  A\&A, 537, A37

\bibitem[{Patil {et~al}\mbox{.}(2010)Patil, Huard, \& Fonnesbeck}]{phf10}
Patil A., Huard D., Fonnesbeck C.~J., 2010, Journal of Statistical Software,
  35, 1

\bibitem[{Phillips \& Wolszczan(1992)}]{pw92}
Phillips J.~A., Wolszczan A., 1992, ApJ, 385, 273

\bibitem[{Press {et~al}\mbox{.}(2007)Press, Teukolsky, Vetterling, \&
  Flannery}]{ptvf07}
Press W.~H., Teukolsky S.~A., Vetterling W.~T., Flannery B.~P., 2007, Numerical
  Recipes: {T}he Art of Scientific Computing, 3$^{rd}$ edition. Cambridge
  University Press, Cambridge

\bibitem[{Scargle(1982)}]{sca82}
Scargle J.~D., 1982, ApJ, 263, 835

\bibitem[{Stairs {et~al}\mbox{.}(2001)Stairs, Manchester, Lyne, Kaspi, Camilo,
  Bell, D'Amico, Kramer, Crawford, Morris, McKay, Lumsden, Tacconi-Garman,
  Cannon, Hambly, \& Wood}]{sml+01}
Stairs I.~H. {et~al.}, 2001, MNRAS, 325, 979

\bibitem[{Standish(1998)}]{sta98b}
Standish E.~M., 1998, JPL Planetary and Lunar Ephemerides, DE405/LE405, Memo
  IOM 312.F-98-048. JPL, Pasadena,
  http://ssd.jpl.nasa.gov/iau-comm4/de405iom/de405iom.pdf

\bibitem[{Stappers {et~al}\mbox{.}(2003)Stappers, Gaensler, Kaspi, van~der
  Klis, \& Lewin}]{sgk+03}
Stappers B.~W., Gaensler B.~M., Kaspi V.~M., van~der Klis M., Lewin W. H.~G.,
  2003, Science, 299, 1372

\bibitem[{{Tam} {et~al}\mbox{.}(2010){Tam}, {Stairs}, {Wagner}, {Kramer},
  {Manchester}, {Lyne}, {Camilo}, \& {D'Amico}}]{tsw+10}
{Tam} C.~R., {Stairs} I.~H., {Wagner} S., {Kramer} M., {Manchester} R.~N.,
  {Lyne} A.~G., {Camilo} F., {D'Amico} N., 2010, MNRAS, 406, 1848

\bibitem[{Tavani {et~al}\mbox{.}(1994)Tavani, Arons, \& Kaspi}]{tak94}
Tavani M., Arons J., Kaspi V.~M., 1994, ApJ, 433, L37

\bibitem[{Vink {et~al}\mbox{.}(2000)Vink, de~Koter, \& Lamers}]{vdl00}
Vink J.~S., de~Koter A., Lamers H.~J.~G.~L.~M., 2000, A\&A, 362, 295

\bibitem[{{Vo{\^ u}te} {et~al}\mbox{.}(2002){Vo{\^ u}te}, {Kouwenhoven}, {van
  Haren}, {Langerak}, {Stappers}, {Driesens}, {Ramachandran}, \&
  {Beijaard}}]{vkv02}
{Vo{\^ u}te} J.~L.~L., {Kouwenhoven} M.~L.~A., {van Haren} P.~C., {Langerak}
  J.~J., {Stappers} B.~W., {Driesens} D., {Ramachandran} R., {Beijaard} T.~D.,
  2002, A\&A, 385, 733

\bibitem[{Wex(1998)}]{wex98}
Wex N., 1998, MNRAS, 298, 67

\end{thebibliography}
\bibliographystyle{mn2e}


\bsp 

\label{lastpage}

\end{document}